\documentclass[aps,prd,twocolumn,
               notitlepage,mathrsfs,
               amssymb,amsmath,amsfonts,
               superscriptaddress,
               longbibliography,
               nofootinbib,floatfix]{revtex4-2}
              
\usepackage[normalem]{ulem}
\usepackage{graphicx}
\usepackage[linktocpage,breaklinks]{hyperref}
\usepackage[capitalize]{cleveref}
\usepackage[usenames,dvipsnames]{xcolor}
\hypersetup{colorlinks=true,
            citecolor=NavyBlue,
            linkcolor=magenta,
            urlcolor=magenta}
\usepackage{multirow,array}
\DeclareMathAlphabet{\pazocal}{OMS}{zplm}{m}{n}
\usepackage{journals}
\usepackage{subcaption}
\usepackage{fmtcount} % equivalent to \usepackage[super]{nth}
\usepackage{upgreek}

\newcommand{\nn}{\nonumber\\}
\newcommand{\Rs}{R_\star}
\newcommand{\pa}{\partial}

\begin{document}
%%%%%%%%%%%%%%%%%%%%%%%%%%%%%%%%%%
% % %
\title{Axisymmetric stability of neutron stars as extreme rotators in massive scalar-tensor theory}

\date{\today}
%%  Add Abbreviations for Journals

\author{Alan Tsz-Lok Lam}
\email{tszlok.lam@aei.mpg.de}
\affiliation{Max Planck Institute for Gravitational Physics (Albert Einstein Institute), 14476 Potsdam, Germany}
	
\author{Kalin V. Staykov}
\affiliation{Department of Theoretical Physics, Faculty of Physics, Sofia University, Sofia 1164, Bulgaria}

\author{Hao-Jui Kuan}
%\email{hao-jui.kuan@aei.mpg.de}
\affiliation{Max Planck Institute for Gravitational Physics (Albert Einstein Institute), 14476 Potsdam, Germany}

\author{Daniela D. Doneva}
%\email{daniela.doneva@uni-tuebingen.de}
\affiliation{Theoretical Astrophysics, Eberhard Karls University of T\"ubingen, T\"ubingen 72076, Germany}
	
\author{Stoytcho S. Yazadjiev}
%\email{yazad@phys.uni-sofia.bg}
\affiliation{Theoretical Astrophysics, Eberhard Karls University of T\"ubingen, T\"ubingen 72076, Germany}
\affiliation{Department of Theoretical Physics, Faculty of Physics, Sofia University, Sofia 1164, Bulgaria}
\affiliation{Institute of Mathematics and Informatics, 	Bulgarian Academy of Sciences, 	Acad. G. Bonchev St. 8, Sofia 1113, Bulgaria}

\begin{abstract}
Differentially rotating scalarized neutron stars, mimickers of binary merger remnants, can possess an enormous angular momentum larger than what could possibly be sustained in a neutron star in general relativity by about one order of magnitude. A natural question to ask is whether these solutions are stable and thus can realize in a binary coalescence. With this motivation in mind, we examine the criterion of dynamical stability against axisymmetric perturbations for these ultra-rotators by numerically tracking their nonlinear evolution in an axisymmetric setup. We demonstrate that the turning-point criterion still serves as a sufficient condition for asymmetric (in)stability. Our findings open an interesting question of whether the merger of two scalarized neutron stars can produce (possibly short-lived) ultra-highly rotating merger remnants.

\end{abstract}
\maketitle

%%%%%%%%%%%%%%%%%%%%%%%%%%%%%%%%%%
\section{Introduction}
%%%%%%%%%%%%%%%%%%%%%%%%%%%%%%%%%%

In the next (\ordinalnum{5}; expected to start in 2027) observing of the international gravitational wave (GW) network, more binary neutron star (BNS) mergers are expected to be witnessed.
The improving sensitivity of the observatory, especially the high-frequency band $\gtrsim10^{3}$~Hz, is thought promising to further resolve the waveforms produced in the post-merger phases. Although certain important information can already be acquired from the pre-merger waveforms such as the bulk properties of the source and the adiabatic tidal response of the neutron star (NS) members \cite{abbo18a,abbo18b}, the post-merger segment of the waveforms delivers information supplementary to the aforementioned ones \cite{shib05b,rezz16,soul22,wijn22}. In particular, the newly formed hypermassive NS (HMNS), which is supported against radial collapse by differential rotation, thermal pressure, and/or magnetic force, carries rich information. For example, the oscillations frequencies and the lifetime of HMNS are not only strongly tied to the internal structure of the star (i.e., to the nuclear equation of state (EOS); e.g., \cite{hoto11,hoto13,Stergioulas:2011gd,Bauswein:2012ya,Bauswein:2015vxa,Rezzolla:2016nxn}), but closely related to the nature of gravity (e.g., \cite{Lam:2024wpq,Lam:2024azd}). 

Scalar-tensor theories of gravity are amongst the most natural and well-motivated alternatives to general relativity (GR). 
Considering the Damour-Esposito-Farese type of scalar-tensor theory (DEF theory hereafter), current pulsar timing observations severely constrain the massless scalar field sector of the theory \cite{zhao22} while only weak bounds can be imposed in the massive case \cite{alsi12}, namely a lower bound of $m_\phi \gtrsim 10^{-15}$~eV \cite{rama16,yaza16} on the scalar mass. 
The constraint on $m_\phi$ can be pushed further by the null evidence of scalarization in the detected waveform ($\lesssim500$~Hz) of GW170817 \cite{abbo19,abbo19prx}. 
In particular, the progenitors of GW170817 are unlikely to be scalarized if the scalar field is massless \cite{zhao19,meht23}, while a scalarized progenitor can still be reconciled with the observed waveform if the scalar field is sufficiently massive with $> 10^{-12}$~eV \cite{kuan23,Xie:2024xex}.
On top of binary systems dynamics, the x-ray pulse profiles emitted by hot spots at NS surfaces infer the mass and radius of NSs \cite{watt16}, which in turn can be used as an independent probe to the EOS and gravitational nature \cite{sota17,silv19,silv19b,hu21,tuna22}.

Within the valid parameter space, Refs.~\cite{done13,done16,done18,stay23} demonstrated in a series of works the existence of stationary, axisymmetric scalarized NSs with an angular momentum exceeding the maximum in GR for a given EOS and rotational law. 
Such super-rotating NSs have very similar properties to the HMNSs produced after mergers of BNS, which inherit most of the angular momentum of the progenitor binary, thus spinning differentially at a large rate. 
Although the maximal angular momentum of HMNSs produced by the merger of non-spinning, quasi-circular binaries is roughly bounded as $J\lesssim8$ in GR (e.g, \cite{uryu09,tani10,tich12}), larger values may be achievable in mergers of dynamically-formed binaries in globular clusters or mergers of NSs having high spins.

The determination of stability of these super-rotating scalarized NSs can be expected to limit the class of HMNSs in the post-merger phase.
The turning-point criterion has been shown to be powerful in detecting secular instabilities and in most cases, its results coincide with the ones from perturbation analysis, i.e., the onset of instability is typically associated with an extremum of a properly chosen function of the stellar equilibrium properties \cite{sork81,sork82}.
The onset location of secular instability has also been studied by numerical simulations \cite{cook92,cook94a,cook94b,shib00a,font02,baio04}, where the validity of the turning-point criterion for uniformly rotating NSs is confirmed.

Here, we briefly recap the turning-point criterion. For axisymmetric spinning NSs and assuming a barotropic EOS, the gravitational mass $M_G$ of an NS can be parameterized by the central energy density $\epsilon_c$, i.e., $M_G=M_G(\epsilon_c)$. In addition, the variations in $M_G$, angular momentum $J$, and baryon number $N$ are related via
\begin{align}
    dM_G=\Omega dJ + \mu dN,
\end{align}
where $\Omega$ is the angular velocity of the star, and $\mu$ is the chemical potential.
The turning-point theorem states that the point where $dJ/d\epsilon_c=dN/d\epsilon_c=0$ separates the stable segment from the secularly unstable one, and the segment satisfying
\begin{align}
    \frac{d\Omega}{d\epsilon_c}\frac{dJ}{d\epsilon_c}+\frac{d\mu}{d\epsilon_c}\frac{dN}{d\epsilon_c}>0
\end{align}
is on the unstable side \cite{frie88}.
From this, we see that the onset of secular instability is marked by the turn-point of $M_G$ along a one-parameter curve with a fixed $J$ or total baryon mass $M_0$.
In particular, for a sequence of equilibria with a fixed $J$, the tuning-point,
\begin{align}\label{eq:turn1}
    \frac{\pa M_G(\epsilon_c)}{\pa \epsilon_c}\bigg\vert_{J}=0
\end{align} 
corresponds to the configuration having the maximal $N$ and thus $M_0$, while the turning-point for a constant $M_0$ sequence, i.e.,
\begin{align}
    \frac{\pa M_G(\epsilon_c)}{\pa \epsilon_c}\bigg\vert_{M_0}=0,
\end{align} 
reflects the minimum of $J$ \cite{cook92}. 
A remark to be made is that when deriving the theorem for uniformly rotating equilibria, Friedman \textit{et al.} \cite{frie88} assumed that, due to viscosity, uniformly rotating equilibria will never become differentially rotating as the final state, i.e., the rotational law can be maintained after the perturbation is damped by viscosity. In other words, this criterion is established by comparing neighboring, rigidly rotating configurations along the one-parameter curve.

On the other hand, the applicability of this theorem to NSs obeying a differential rotation law is not established analytically since the rotational law may be altered by any perturbation. Thus, the equilibria do not form a one-parameter family but rather a family of infinite dimensions (to which a turning-point theorem is still possible to hold in some form \cite{sork82}).
While not shown analytically, numerical studies suggest that the turning-point criterion approximately applies to differential rotating NSs \cite{kapl14,weih18,muha24} in GR.
The goal here is to examine the validity of the criterion for high-$J$ stars in scalar-tensor theories, which have no counterparts in GR, through axisymmetric numerical relativity simulations. 
In particular, we will numerically evolve the stellar profiles to determine if the configuration is stable to a random numerical perturbation, or if some instabilities will operate so that the initial state will migrate to a final state which may be another neutron star configuration or a black hole. 

This short paper is organised as follows: \cref{formalism} introduces the basic equations for constructing differentially rotating, scalarized NSs and the scheme for axisymmetric evolution. We provide the numerical results in \cref{results} and discuss them in \cref{discussion}.
Unless specified otherwise, we adopt the geometric unit of $G=c=1$ throughout this paper, where $G$ and $c$ are the gravitational constant and speed of light, respectively. 
The subscripts $a$, $b$, $c, \cdots$ denote the spacetime coordinates while $i$, $j$, $k, \cdots$ the spatial coordinates, respectively. 

%%%%%%%%%%%%%%%%%%%%%%%%%%%%%%%%%%
\section{Basic Equations}\label{formalism}
%%%%%%%%%%%%%%%%%%%%%%%%%%%%%%%%%%

The Jordan frame action of the DEF theory is given by \cite{damo92}
\begin{align}\label{eq:action}
    S =& \frac{1}{16\pi}\int d^4x \sqrt{-g}
    \left[ \phi {\cal R} - \frac{\omega(\phi)}{\phi}
    \nabla_a \phi \nabla^a \phi - U(\phi) \right] \nonumber\\
    &- S_{\rm matter},
\end{align}
where $\mathcal{R}$ and $g$ are, respectively, the Ricci scalar and the determinant of the metric function $g_{ab}$, $\nabla_a$ is the covariant derivative associated with $g_{ab}$, $\phi$ is the scalar field, and $S_{\rm matter}$ is the action of matter. 
The coupling function, $\omega(\phi)$, is expressed as \cite{shib14,tani15,senn16}
\begin{align}
    \frac{1}{\omega(\phi)+3/2}=B\ln\phi,
\end{align}
with $B$ being a dimensionless free parameter, and the potential of the scalar field \cite{kuro23,kuan23},
\begin{align}
    U(\phi) = \frac{2m_\phi^2\varphi^2\phi^2}{B},
\end{align}
gives rise to a mass term $m_\phi$ for the scalar field.
In the above expression, we have defined $\varphi=\sqrt{2\ln\phi}$. 
When transformed into the mathematically convenient Einstein frame by rephrasing the action \eqref{eq:action} in terms of an auxiliary metric $g_{ab}^E=\phi\, g_{ab}$, this potential can be shown to have the form of $V=m_\phi^2\bar\varphi^2/(2B)$ with $\bar\varphi=\varphi/\sqrt{B}$. The meaning of scalar mass then becomes clear.
In a portion of the literature, the factor between $g_{ab}$ and $g_{ab}^E$ is referred to as the coupling function.
Under this context, the conventional expression for the coupling function reads $\phi=e^{-\beta \bar\varphi^2}$ for $\beta=-B/2$.
The formulation for constructing initial data in the considered theory will be described in \cref{rns}.
The detailed setup of numerical evolution will then follow in \cref{sacra2d}.

%%%%%%%%%%%%%%%%%%%%%%%%%%%%%%%%%%
\subsection{Profiles from RNS}\label{rns}
%%%%%%%%%%%%%%%%%%%%%%%%%%%%%%%%%%

A modified \texttt{RNS} code \cite{ster95} for generating initial data of equilibrium states of scalarized NSs in the DEF theory has been developed in a series of works \cite{done13,done18,stay23} from simpler to more sophisticated rotation laws.  
The code uses a modified \cite{cook92} Komatsu--Eriguchi--Hachisu (KEH) \cite{koma89} scheme for constructing rotating equilibrium neutron star models. For mathematical and numerical convenience, the calculation of equilibrium models is performed in the so-called Einstein frame, which is later transformed into the physical Jordan frame used by the evolution code. 
The two frames are related through a conformal transformation of the metric, and a detailed discussion can be found in \cite{done13}.

The modified \texttt{RNS} code adopts quasi-isotropic coordinate, in which the metric is expressed in the spherical coordinate $(r, \theta, \upphi)$ as \cite{evan84,koma89,koma89b}
\begin{align}
    ds^2 &= -e^{\eta+\sigma}dt^2 
    + e^{\eta-\sigma}r^2\sin^2\theta (d\upphi^2-\varpi dt)^2 \nn
    &+ e^{2\tau}(dr^2 + r^2d\theta^2) \\
    =& -(\alpha^2-\gamma_{\upphi\upphi}\varpi^2)dt^2
    -2\varpi \gamma_{\upphi\upphi}d\upphi dt + \gamma_{ij}dx^{i}dx^{j},
    \nonumber
\end{align}
where $\alpha$ and $\gamma_{ij}$ is the lapse function and spatial metric, respectively.
The shift vector $\beta^i$ is expressed as  (see, e.g., Sec.~4 of \cite{pasc17})
\begin{align}\label{eq:shift}
 \beta^i=-\varpi(\partial_\upphi)^i.
\end{align}
Here, $\varpi$ is the frame-dragging factor, and the spatial metric $\gamma_{ij}$ is
\begin{align}
\begin{split}
    \gamma_{ij}&=\psi^4
    \begin{pmatrix}
        e^{-\mathbf{q}} & 0 & 0 \\
        0 & e^{-\mathbf{q}} r^2 & 0 \\
        0 & 0 & e^{2\mathbf{q}}  r^2\sin^2\theta
    \end{pmatrix} \\
    &= \psi^4 \tilde \gamma_{ij}
\end{split}
\end{align}
for
\begin{align}
    \psi = e^{(4\tau + \eta - \sigma)/12} \quad\text{and}\quad \mathbf{q} = \frac{2}{3} \left( 2 \tau - \eta + \sigma \right),
\end{align}
%---------------------------
where $\psi$ is the conformal factor and $\tilde \gamma_{ij}$ is the conformal spatial metric with its determinant $\det (\tilde \gamma_{ij}) = \det (f_{ij})$ same as the flat background metric $f_{ij}$.
In the above expressions, $\tau$, $\varpi$, $\sigma$, and $\eta$ are all functions of $r$ and $\theta$ only since we consider axisymmetric NSs. 
For non-spinning NSs, we have $e^{\eta-\sigma}=e^{2\tau}$, and thus the metric $\gamma_{ij} = e^{2\tau} f_{ij}$ is conformally flat, while the metric will be distorted from the conformal flatness due to the dragging effect when the star is rotating.
The determinant of $\gamma_{ij}$ in the Cartesian coordinates is $\gamma:=\det{(\gamma_{ij})} = e^{4\tau+\eta-\sigma}$, which again reduces to $\gamma=e^{6\tau}$ for non-rotating configurations.
For the considered gauge and coordinate, the extrinsic curvature tensor, defined as
\begin{align}
    2\alpha K_{ij} = D_i\beta_j+D_j\beta_i,
\end{align}
has the form (see, e.g., Eqs.~(2.43) and (2.44) of \cite{gour10})
\begin{align}
    K_{ij}=\frac{-e^{\eta}r^2\sin^2\theta}{2 \alpha}
\begin{pmatrix}
    0 & 0 & \frac{\partial}{\partial r} \varpi \\
    \cdot & 0 & \frac{\partial}{\partial\theta} \varpi \\
    \cdot & \cdot & 0
\end{pmatrix},
\end{align}
where ``$\cdot$''s are the ellipsis of the symmetric part
and $D_i$ is the covariant derivative associated with the spatial metric $\gamma_{ij}$.

For the matter profile, the 4-velocity of matter is expressed as
\begin{align}
    u^{a}=\frac{w}{\sqrt{\alpha^2-\gamma_{\upphi\upphi}\varpi^2}}(1,0,0,\Omega),
\end{align}
with $w:=(1-v^2)^{-1/2}$ being the Lorentz factor and $v$ the proper velocity, given by
\begin{align}
    v=(\Omega-\varpi)r\sin\theta.
\end{align}
The spin of the star,
\begin{align}
    \Omega(r,\theta) =\frac{u^{\upphi}}{u^{t}},
\end{align}
is specified by a certain rotation law as well as the stellar structure. We note that $\Omega$ is the same in both the Einstein and Jordan frames, thus not further complicating the transition of the quantities in the two codes. 

In the present article, we adopt the 4-parameter differential rotation law introduced by Uryu et al. \cite{uryu17,iosi21,iosi22},
\begin{align}\label{eq:rot_law}
    \Omega = \Omega_c \frac{1+\left(\frac{F}{B^2\Omega_c}\right)^p}{1+\left(\frac{F}{A^2\Omega_c}\right)^{p+q}},
\end{align}
where $F=u^tu_\upphi$ is the redshifted angular momentum per unit rest mass. This rotation law allows for the maximum of the angular velocity to be away from the center, which is a common characteristic seen in remnants in merger simulations, e.g., \cite{Kastaun:2014fna, Bauswein:2015yca, DePietri:2019mti}. 
Here, two constants have been fixed to $p=1$ and $q=3$ \cite{zhou19,iosi21,iosi22}. 
This choice allows one to derive an analytical expression for the first integral of the hydrostationary equilibrium, which is required for the \texttt{RNS} code. 
The other two parameters, $A$ and $B$, are not given explicitly. 
Instead, the ratios $\lambda_1 = \Omega_{\max}/\Omega_c$ and $\lambda_2 = \Omega_e/\Omega_c$, where $\Omega_e$ is the angular velocity at the equator, $\Omega_c$ is the angular velocity at the center and $\Omega_{\max}$ is the maximum of the angular velocity, are given. From them, one can obtain and solve an algebraic system for $A$ and $B$. 
Those ratios control the shape of the neutron star. 
In the present article we use $(\lambda_1,\lambda_2) = (1.5,0.5)$ which correspond to the quasi-toroidal models \cite{stay23}. 
When the rotation law $F$ is given, the angular momentum of the star is determined via
\begin{align}
    J=\int_{r<\Rs} \alpha\rho hF\sqrt{\gamma}d^3x,
\end{align}
for a given rest-mass density $\rho$ and specific enthalpy $h$ distributions inside the star.

\subsection{Evolution equations}
\label{sacra2d}
The modified evolution equations in DEF theory under Baumgarte-Shapiro-Shibata-Nakamura (BSSN) formulation \cite{shib95,baug98,Lam:2024wpq} in the Cartesian coordinates are written as
\begin{subequations}
\begin{align}
%%%
    ( \pa_t - &\beta^k \pa_k ) W =
    \frac{1}{3} W \left( \alpha K - \pa_k \beta^k \right), \\
%%%
    ( \pa_t - &\beta^k \pa_k ) \tilde \gamma_{ij} = 
    - 2 \alpha \tilde A_{ij} \nonumber \\
    &+ \tilde \gamma_{ik} \pa_j \beta^k
    + \tilde \gamma_{jk} \pa_i \beta^k - \frac{2}{3} \tilde \gamma_{ij} \pa_k \beta^k,
\end{align}
\begin{align}
%%%
    ( \pa_t - &\beta^k \pa_k ) \tilde A_{ij} = 
    W^2 \left[ \alpha R_{ij} -  D_i D_j \alpha - 8\pi \alpha \phi^{-1} S_{ij} \right]^{\rm TF} \nonumber \\
    &+ \alpha \left( K \tilde A_{ij} - 2 \tilde A_{ik} \tilde A_j{}^k \right)
    + \tilde A_{kj} \pa_i \beta^k + \tilde A_{ki} \pa_j \beta^k \nonumber\\
    &- \frac{2}{3} \tilde A_{ij} \pa_k \beta^k + \alpha \tilde A_{ij} \varphi \Phi \nonumber\\
    & - \alpha W^2 \left[ \omega \varphi^2 D_i \varphi D_j \varphi + \phi^{-1} D_i D_j \phi \right]^{\rm TF}, 
%%%
\end{align}
\begin{align}
    (\partial_t - &\beta^k \partial_k) K
    =\,\, 4\pi \alpha \phi^{-1} (S^i{}_i+\rho_{\rm h})+\alpha K_{ij} K^{ij}-D_i D^i \alpha \nonumber\\
    &+\alpha \omega \varphi^{2} \Phi^2 - \left( \frac{3}{2}+\frac{1}{B} \right)\alpha m_\phi^2\varphi^2\phi
    \nonumber \\
    &+ \alpha\phi^{-1}\Big[
    D_iD^i\phi-K\Phi\phi\varphi \nonumber\\
    &-3\pi\varphi^2 BT
    +\frac{3}{2\varphi^2\phi} \left(\Phi^2\phi^2\varphi^2-D_k\phi D^k\phi\right)
    \Big], \label{eq:K} \\
%%%
    (\pa_t - &\beta^k \pa_k ) \tilde \Gamma^i =
    2 \alpha \left( \tilde \Gamma^{i}_{jk} \tilde A^{jk} - \frac{2}{3} \tilde \gamma^{ij} \pa_j K 
    - \frac{3}{W} \tilde A^{ij} \pa_j W \right) \nonumber \\
    & - 2 \tilde A^{ij} \pa_j \alpha - 2\alpha \tilde \gamma^{ij} \left[ 8\pi \phi^{-1} J_j - \varphi K_j{}^k D_k\varphi \nonumber \right. \\
    & \left.+\left(1+ \frac{2}{B}-\frac{\varphi^2}{2} \right)\Phi D_j\varphi
    +\varphi D_j\Phi \right] \nonumber \\
    &+ \tilde \gamma^{jk} \pa_j \pa_k \beta^i + \frac{1}{3} \tilde \gamma^{ij} \pa_j \pa_k \beta^k \nonumber \\
    &- \tilde \gamma^{kl} \tilde \Gamma^j{}_{kl} \pa_j \beta^i + \frac{2}{3} \tilde \gamma^{jk} \tilde\Gamma^i{}_{jk} \pa_l \beta^l, 
\end{align}
\begin{align}
    (\partial_t -&\beta^k \partial_k)\varphi =-\alpha \Phi,\\
%%%
    (\partial_t -&\beta^k \partial_k)\Phi = -\alpha D_i D^i \varphi
    - (D_i \alpha)D^i\varphi
    -\alpha \varphi (\nabla_a \varphi)\nabla^a \varphi \nonumber\\
    &+\alpha K \Phi
    + 2\pi \alpha \phi^{-1} B T \varphi +\alpha m_\phi^2\varphi\phi,
%%%
\end{align}
\end{subequations}
where $W:=\psi^{-2}$,
$\Phi := - n^a \nabla_a \varphi$ is the "momentum" of the scalar field with the time-like unit normal vector $n^a=(1/\alpha, - \beta^i/\alpha)$, 
$\tilde \Gamma^i{}_{jk}$ is the Christoffel symbol associated with $\tilde \gamma_{ij}$, $\tilde \Gamma^i := - \pa_j \tilde\gamma^{ij}$, $(S_{ij})^{\rm TF}:= S_{ij} - \gamma_{ij} S^k{}_k / 3$ denotes the trace-free part of the stress tensor $S_{ij}$, $K:=K^i{}_i$ is the trace of the extrinsic curvature, $\tilde A_{ij} := W^2 (K_{ij})^{\rm TF}$ is the conformal traceless part of $K_{ij}$, 
$R_{ij}$ is the spatial Ricci tensor, 
$T:=T^{a}{}_{a}$ is the trace of the stress-energy tensor, and
$S_{ij}:= \gamma^a{}_i \gamma^b{}_j T_{ab}$, $\rho_h := n^a n^b T_{ab}$ and $J_i := - \gamma^a{}_i n^b T_{ab}$ are the spacetime decompositions of the stress-energy tensor.

We adopted the moving puncture gauge \cite{alcu03,bake06,camp06} for the lapse function and shift vector as
\begin{subequations}
\begin{align}
    ( \partial_t - \beta^j \partial_j ) \alpha &= - 2 \alpha K , \\
    ( \partial_t - \beta^j \partial_j ) \beta^i &= \frac{3}{4} B^i , \\
    ( \partial_t - \beta^j \partial_j ) B^i &= ( \partial_t - \beta^j
    \partial_j ) \tilde{\Gamma}^i - \eta_B B^i,
\end{align}
\end{subequations}
where $B^i$ is an auxiliary variable and $\eta_B$ is a parameter typically set to be $\approx 1/M_G$.

The cartoon method \cite{alcu01} has proven to be a robust scheme to evolve axisymmetric spacetime \cite{shib00b,shib03a,shib03b,shib03c}.
We extended the 2D cartoon code \texttt{SACRA-2D} developed in \cite{lam25} to include the evolution equations of DEF theory with Z4c constraint propagation \cite{bern10,hild13}.
\texttt{SACRA-2D} employs a fixed mesh refinement with 2:1 refinement and imposes equatorial mirror symmetry on the $z = 0$ plane.
For the simulations included in this paper, the differentially rotating NS is covered by 9 refinement levels with at least 150 grid points covering the equatorial radius of the NS.
We adopted \ordinalnum{6} order finite difference for the field equations and HLLC Riemann solver \cite{mign05,whit16,kiuc22} for hydrodynamics.

\begin{table}
    \centering
    \caption{Parameters of different sequences considered in this paper, where the scalar properties (second column), the angular momentum fixed for each sequence (third column), and the central energy density of the NS at the onset of asymmetric instability (last column) are collated.}
    \begin{tabular}{p{0.12\textwidth}|
				>{\centering}p{0.12\textwidth}
                    >{\centering}p{0.08\textwidth}
				>{\centering\arraybackslash}p{0.12\textwidth}}
    \hline
    \hline \\[-.8em]
        Sequence name & ($m_\phi,B$) & $J ~(M_\odot^2)$ & $\epsilon_{\rm thre} ~[\times 10^{15}~\text{cgs}]$ \\   \\[-.8em]      
    \hline\\[-.8em]
        \texttt{m0\_B12\_J8}  & (0, 12) &  8 & 1.16  \\   \\[-.8em]
        \texttt{m0\_B12\_J12} & (0, 12) & 12 & 1.12  \\   \\[-.8em]
        \texttt{m0\_B12\_J20} & (0, 12) & 20 & 1.056 \\   \\[-.8em]
        \texttt{m0\_B12\_J40} & (0, 12) & 40 & NA \\   \\[-.8em]
    \hline\\[-.8em]
        \texttt{m0.01\_B12\_J8}  & (0.01, 12) &  8 & 1.15  \\   \\[-.8em]
        \texttt{m0.01\_B12\_J12} & (0.01, 12) & 12 & 1.08  \\   \\[-.8em]
        \texttt{m0.01\_B12\_J20} & (0.01, 12) & 20 & NA \\   \\[-.8em]
        \texttt{m0.01\_B12\_J30} & (0.01, 12) & 30 & NA \\   \\[-.8em]
    \hline
    \hline
    \end{tabular}
    \label{tab:models}
\end{table}
%%%%%%%%%%%%%%%%%%%%%%%%%%%%%%%%%%
\section{Numerical results}\label{results}
%%%%%%%%%%%%%%%%%%%%%%%%%%%%%%%%%%

\begin{figure*}
    \centering
    \begin{subfigure}[t]{0.48\textwidth}
        \includegraphics[width=\linewidth]{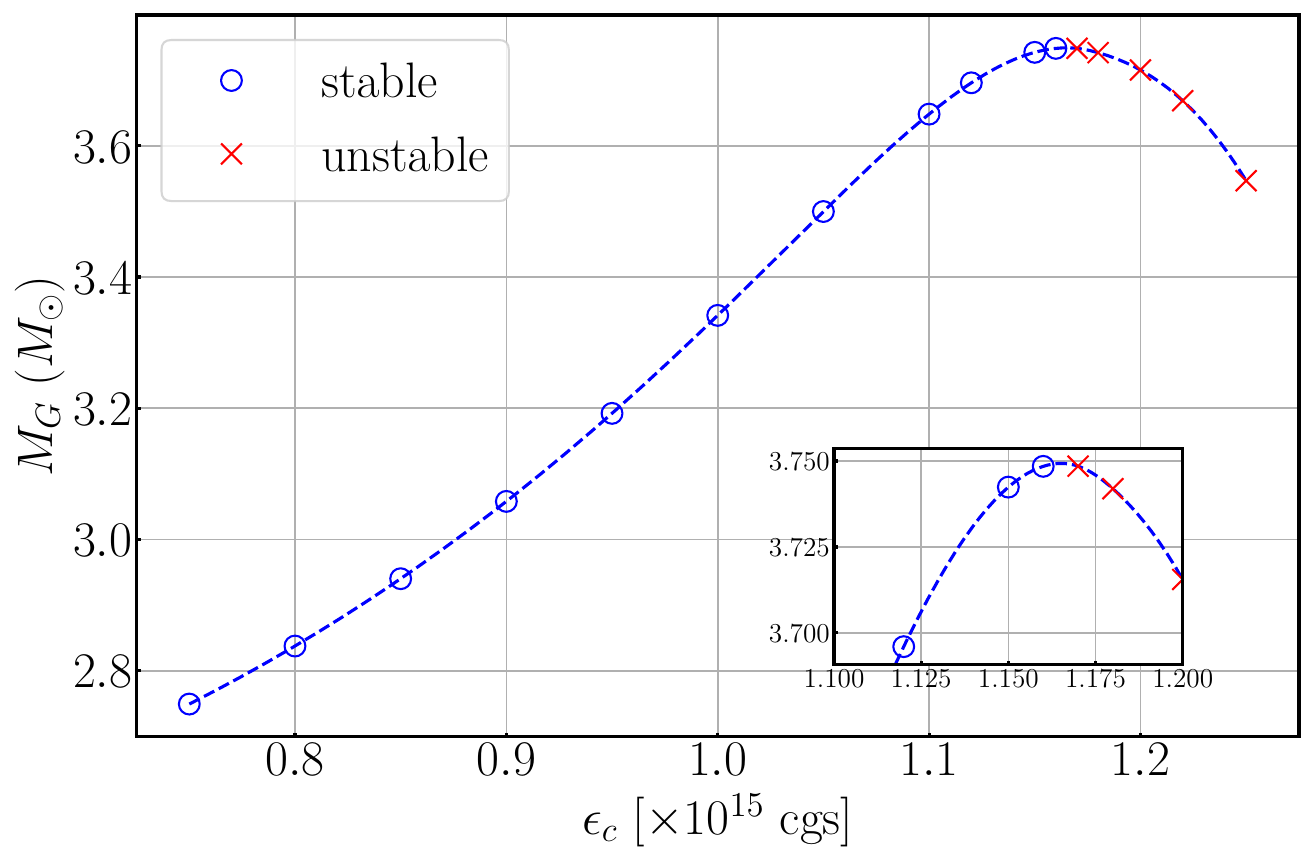}
        \caption{\texttt{m0\_B12\_J8}}
        \label{J8}
    \end{subfigure}
    \hfill %%
    \begin{subfigure}[t]{0.48\textwidth}
        \includegraphics[width=\linewidth]{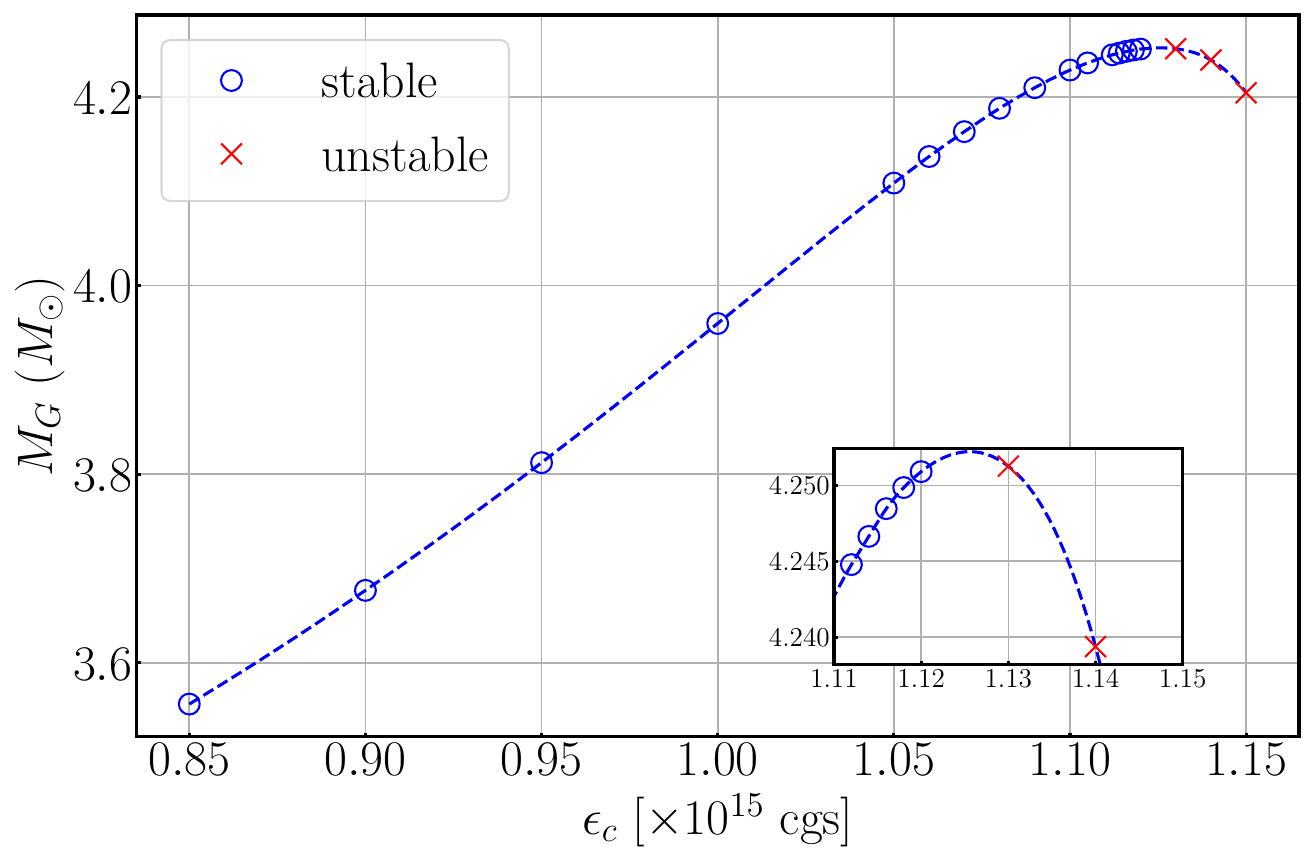}
        \caption{\texttt{m0\_B12\_J12}}
        \label{J12}
    \end{subfigure}
    \vfill
    \begin{subfigure}[t]{0.48\textwidth}
        \includegraphics[width=\linewidth]{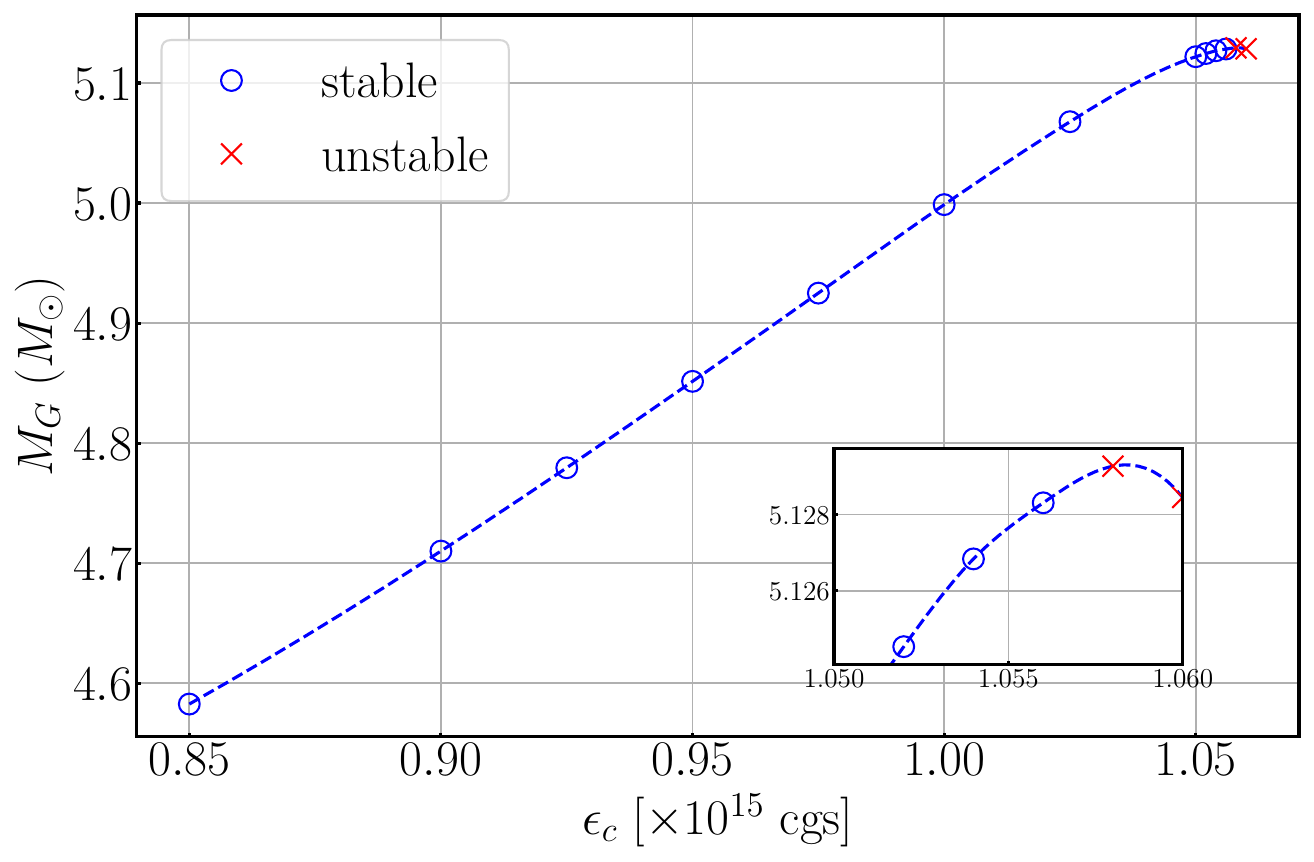}
        \caption{\texttt{m0\_B12\_J20}}
        \label{J20}
    \end{subfigure}
    \hfill %%
    \begin{subfigure}[t]{0.48\textwidth}
        \includegraphics[width=\linewidth]{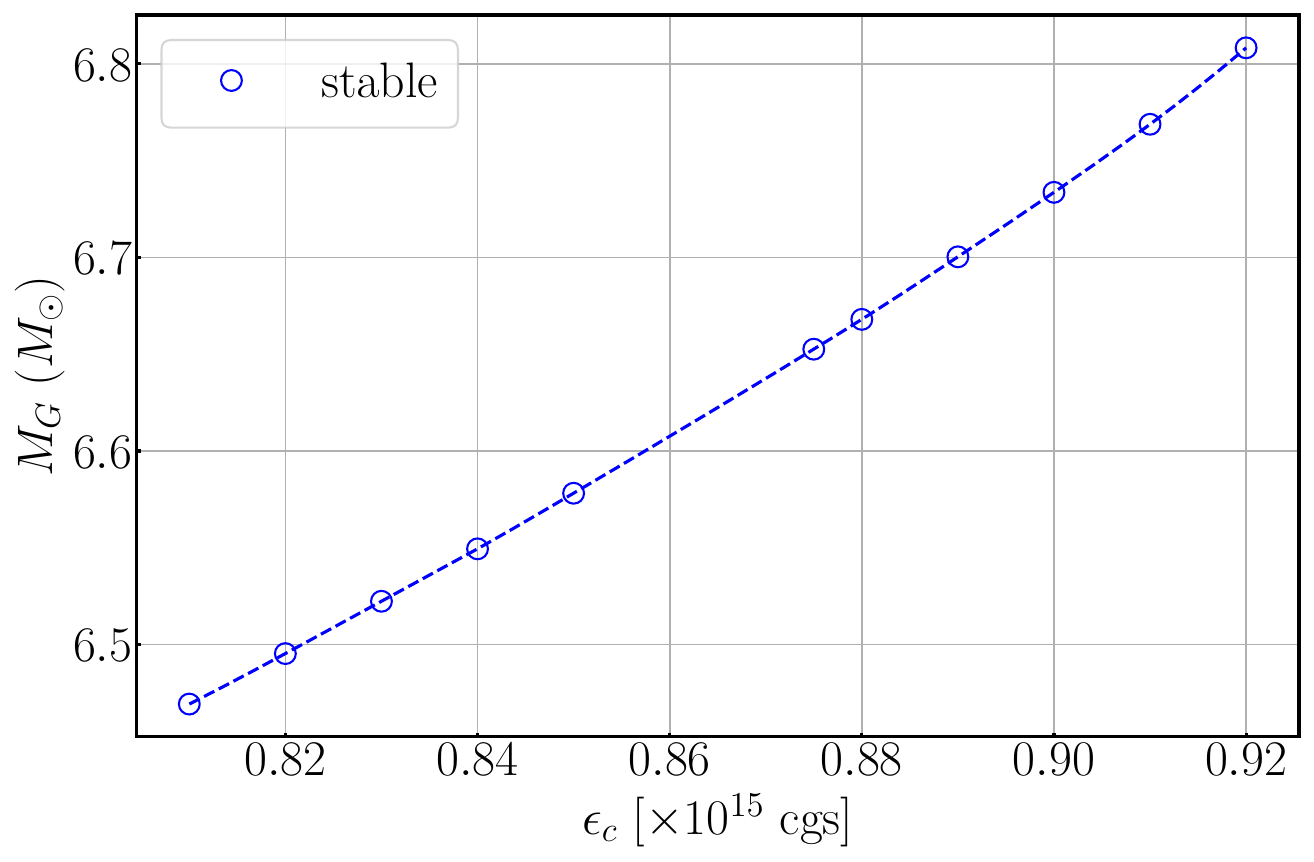}
        \caption{\texttt{m0\_B12\_J40}}
        \label{J40}
    \end{subfigure}
    \caption{Dynamical stability of sequences (a) \texttt{m0\_B12\_J8} (b) \texttt{m0\_B12\_J12} (c) \texttt{m0\_B12\_J20} (d) \texttt{m0\_B12\_J40}. The blue circles and red crosses indicate the regions where the star is dynamically stable and unstable respectively.
    }
    \label{fig:massless}
\end{figure*}
We construct axis-symmetric, spinning NSs obeying the rotational law \eqref{eq:rot_law} for several fixed values of $J$ that will be used as initial data for the nonlinear evolution code. The representative sequences are summarized in \cref{tab:models}, where we consider the massless DEF theory and a massive scalar field theory with $m_\phi=0.01$ ($\simeq 1.33\times10^{-12}$~eV). We fix $B=12$ and focus on MPA1 EOS \cite{muth87} as a representative example. 
We perform axis-symmetric relativistic simulations for selected models from these sequences, especially close to the maximum mass point, where possible. The goal is to examine the stability and study the outcome of unstable models. We start with the massless theory (\cref{sec.III.A}) followed by a study of the massive scalar field case (\cref{sec.III.B}).

\subsection{Massless scalar field case}\label{sec.III.A}
For the massless cases, a turning point can be found for three of the considered angular momenta in Table \ref{tab:models}, while the sequence with $J=40$ has no turning point, i.e. no maximum of the mass was reached. 
A general behavior of the solutions generated by the \texttt{RNS} code is that with the increase of the angular momentum, the solution branches get shorter, and they get terminated before reaching the turning point. 
The reason is numerical -- the \texttt{RNS} code can not converge to a unique solution. 
Different numerical schemes may be useful to overcome this problem, such as the spectral method used in \cite{Ansorg:2008pk,Gondek-Rosinska:2016tzy}, that is out of the scope of the present paper.

\begin{figure}
    \centering
    \includegraphics[width=.9\linewidth]{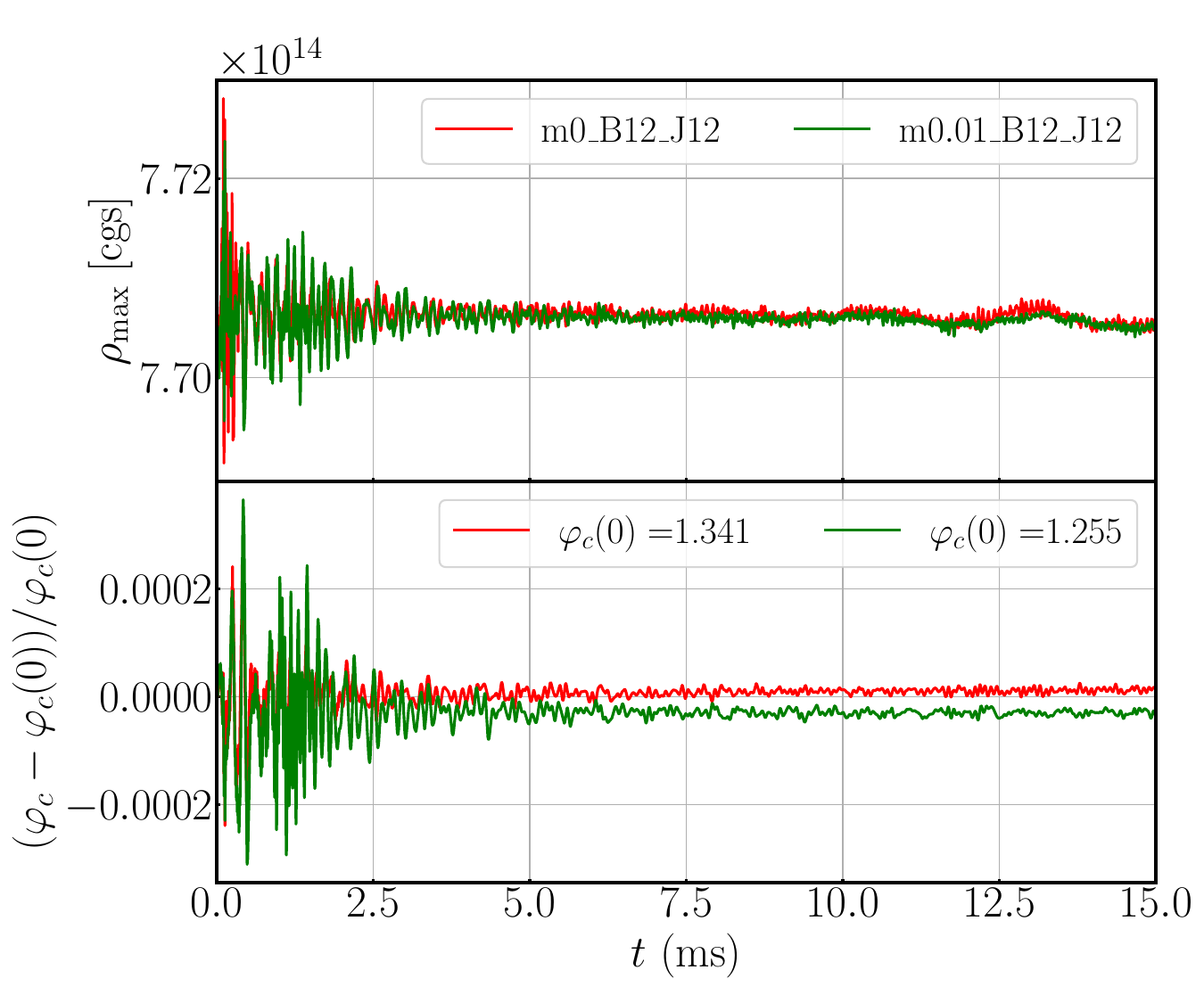}
    \caption{Evolution of maximum density $\rho_{\rm max}$ (top) and central scalar field $\varphi_{c}$ (bottom) for two stable models with an initial central energy density $\epsilon_c = 8.5 \times 10^{14}$~${\rm g}/{\rm cm}^3$.}
    \label{fig:stable_evo}
\end{figure}

\begin{figure}
    \centering
    \includegraphics[width=.9\linewidth]{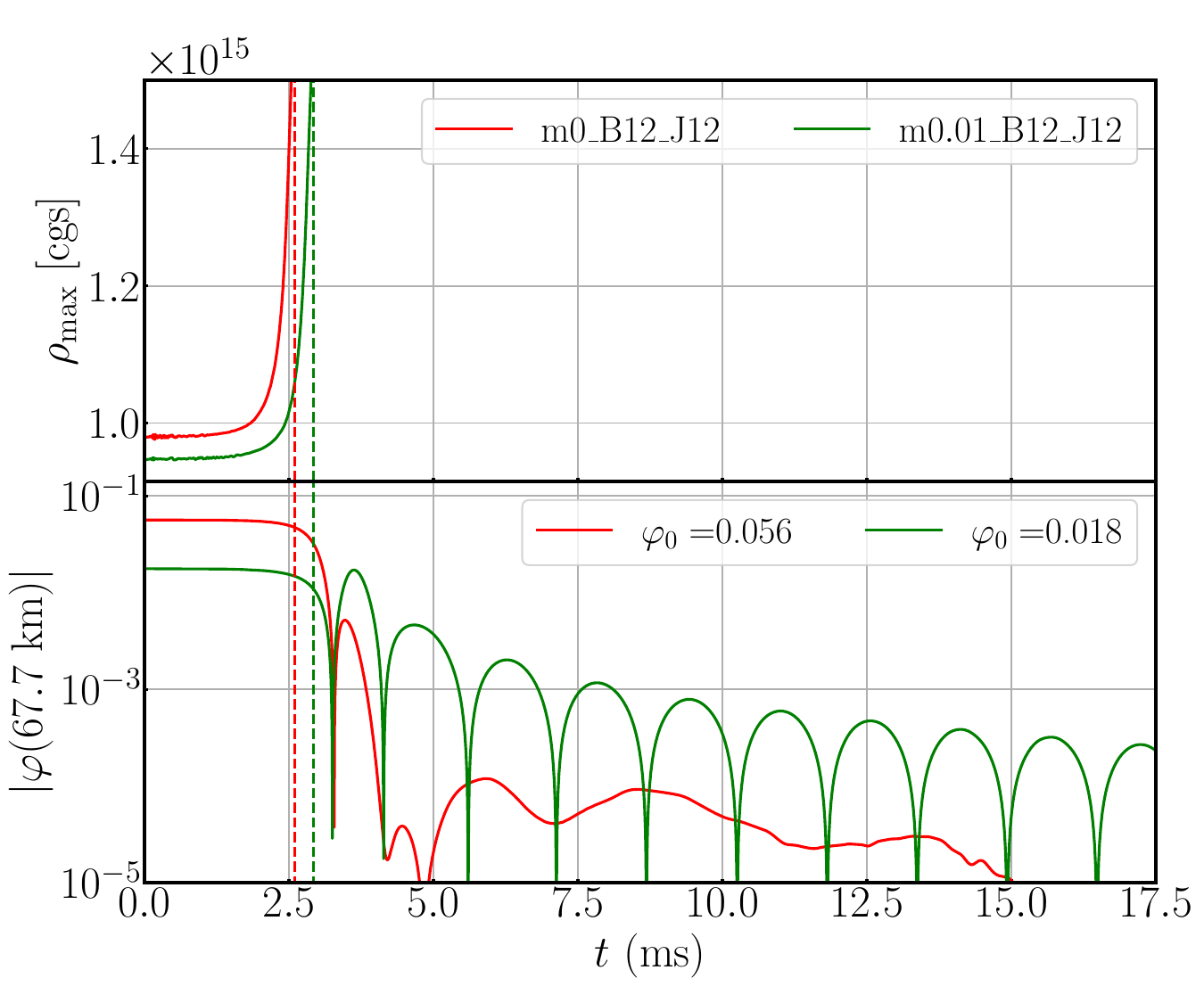}
    \caption{Evolution of maximum density $\rho_{\rm max}$ (top) and absolute value of central scalar field $|\varphi_{c}|$ (bottom) for unstable models with initial central energy density $\epsilon_c = 1.15 \times 10^{15}$ ${\rm g}/{\rm cm}^3$ (red) and $\epsilon_c = 1.10 \times 10^{15}$ ${\rm g}/{\rm cm}^3$ (green).
    The colored dashed lines represent the formation time of an apparent horizon for the corresponding models.}
    \label{fig:unstable_evo}
\end{figure}

Along each sequence, we study the (asymmetric) stability of 10--20 models, most of which condense near the maximum mass point. 
The results are summarized in \cref{fig:massless}, where we see that for cases a), b) and c) the marginally stable model is slightly left to the maximum of the mass, implying that the turning-point criterion approximately predicts the onset of instability for these sequences. For case d), where no turning point was reached in the equilibrium sequence, all neutron star models are stable.
For the stable models, the perturbations in the maximum density and central scalar field damp in a dynamical timescale, then settle back to the initial values. Taking the model \texttt{m0\_B12\_J20} as an example, which is very close to the turning-point along the sequence of $J=20$, the evolution of the maximum rest-mass density and the central value of the scalar field are shown \cref{fig:stable_evo} (red), where we see that the initial noise is dissipated after $<5$~ms.
On the other hand, the unstable models will collapse into a black hole in a dynamical timescale. 
For one such example \texttt{m0\_B12\_J12}, the evolution of the maximal rest-mass density shows a runaway growth in less than 3~ms (red in \cref{fig:unstable_evo}).
After the formation of a black hole, the scalar field dissipates exponentially to $<10^{-4}$ since black holes in this theory obey the no-hair theorem and thus cannot possess a stationary scalar field \cite{hawk72,soti12}.

\begin{figure*}
    \centering
    \includegraphics[width=\linewidth]{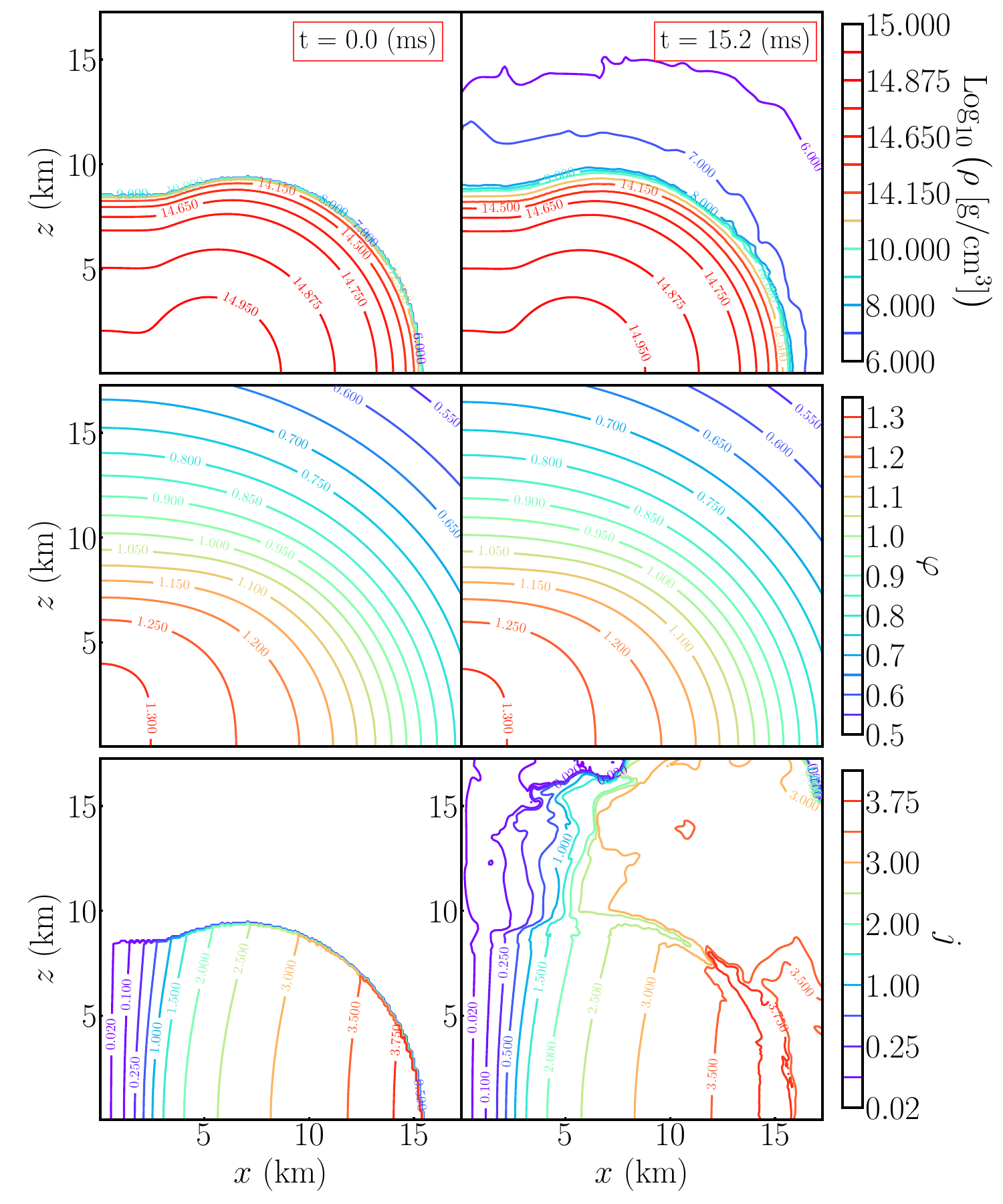}
    \caption{Snapshots for a stable model in \texttt{m0\_B12\_J20}, whose central energy density is $\epsilon_c = 1.052 \times 10^{15}$~${\rm g}/{\rm cm}^3$. The initial profiles are shown in the left column, including rest-mass density (top), scalar field (middle), and specific angular momentum in the code unit (bottom). The profiles at 15.2 ms for them are shown in the right column respectively.}
    \label{fig:snapshot}
\end{figure*}

In addition, the rotational law is well-maintained over several dynamical timescales in our simulations for stable models.
For one stable example, we plot the profiles of rest-mass density, scalar field, and the specific angular momentum,
\begin{align}
    j:=hu_\upphi,
\end{align}
at the initial moment and at 15~ms in \cref{fig:snapshot}.
Apart from a tiny amount of matter that escapes from the surface, the structure of rest-mass density and scalar field remain unaltered to a large extent, i.e., the model is rather stable to axisymmetric perturbations, and the numerical accuracy is robust.
In particular, the profile of specific angular momentum within the HMNS is well preserved after $15$ ms, showing that the rotational profile is also stable under such perturbations.
We also note that the onset of instability is not sensitive to the employed resolutions.

\subsection{Massive case}\label{sec.III.B}
We also examine the criterion along fixed-$J$ sequences for a massive scalar field with $m_\phi=0.01$($\simeq 1.33 \times10^{-12}$~eV). This value is chosen in order to be in agreement with binary neutron star merger observations \cite{kuan23}. It is a rather large value, and it effectively confines the scalar field in a radius several times larger with respect to the neutron star size.

\begin{figure*}
    \centering
    \begin{subfigure}[t]{0.48\textwidth}
        \includegraphics[width=\linewidth]{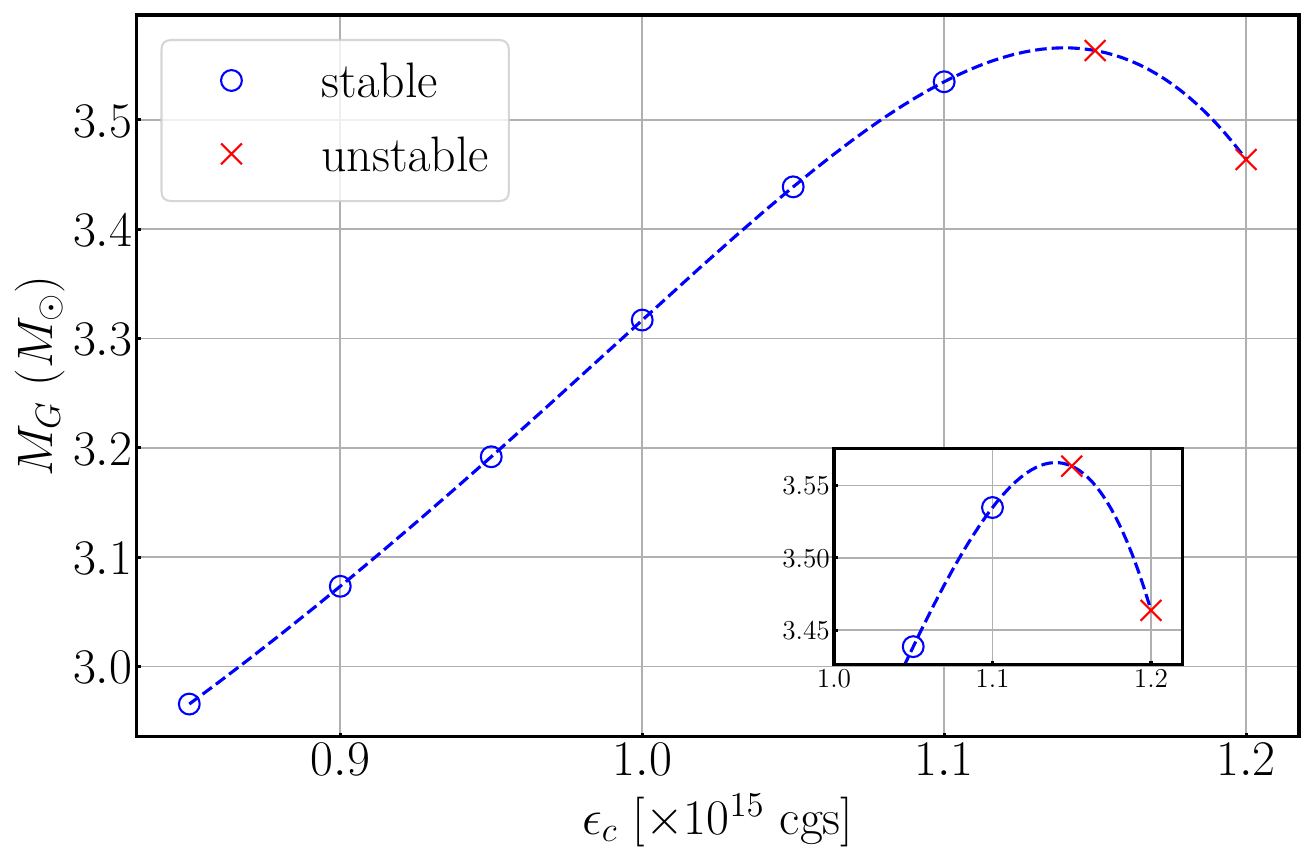}
        \caption{\texttt{m0.01\_B12\_J8}}
        \label{m0.01_J8}
    \end{subfigure}
    \hfill %%
    \begin{subfigure}[t]{0.48\textwidth}
        \includegraphics[width=\linewidth]{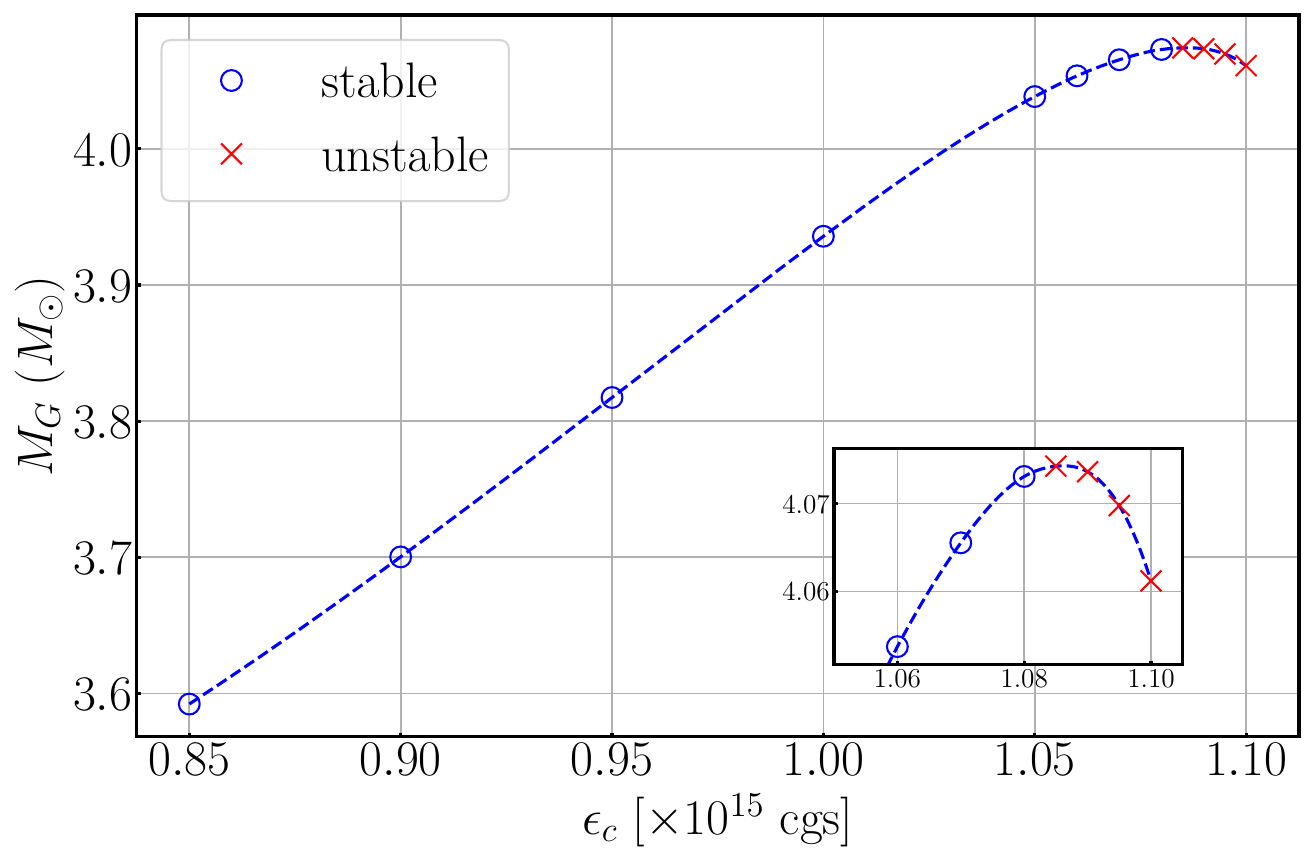}
        \caption{\texttt{m0.01\_B12\_J12}}
        \label{m0.01_J12}
    \end{subfigure}
    \vfill
    \begin{subfigure}[t]{0.48\textwidth}
        \includegraphics[width=\linewidth]{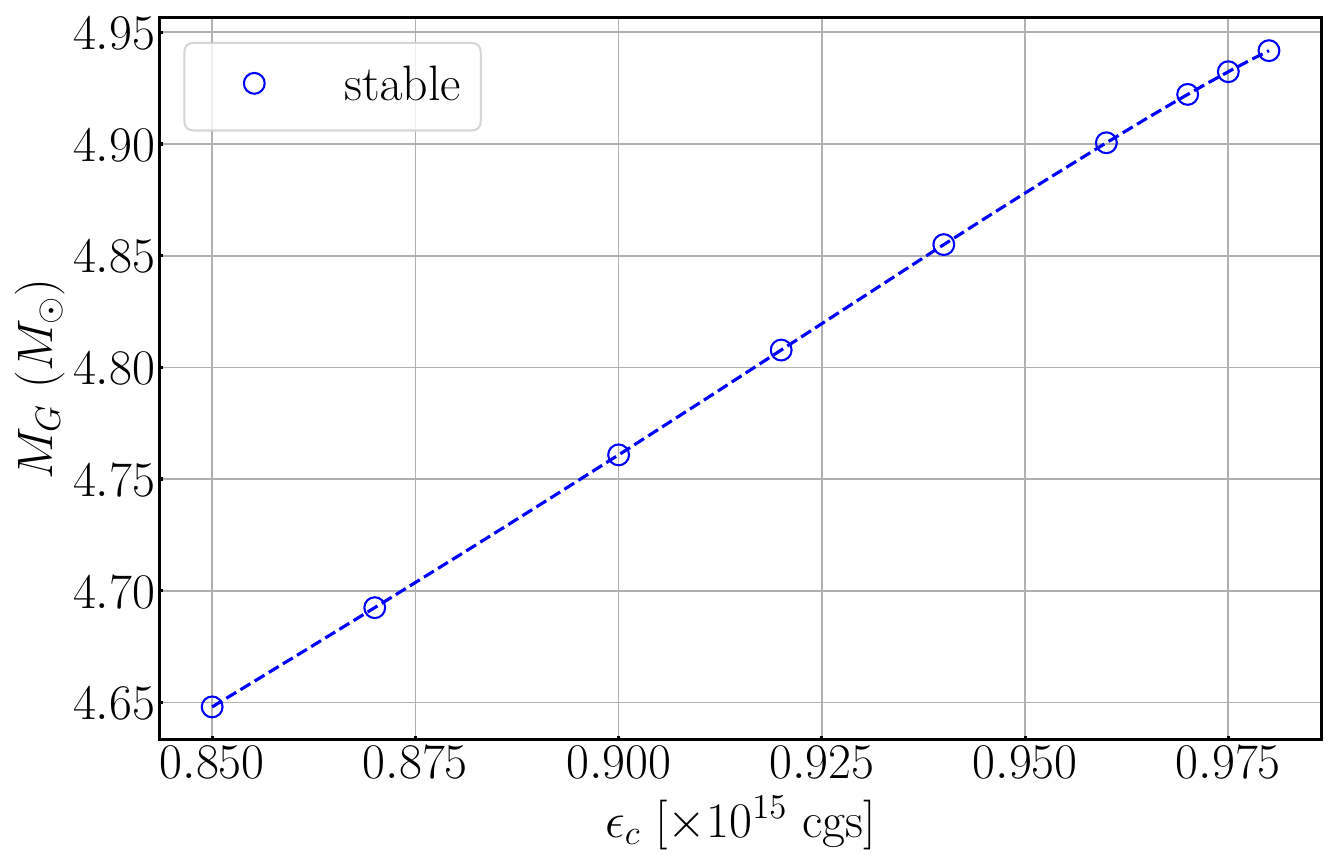}
        \caption{\texttt{m0.01\_B12\_J20}}
        \label{m0.01_J20}
    \end{subfigure}
    \hfill %%
    \begin{subfigure}[t]{0.48\textwidth}
        \includegraphics[width=\linewidth]{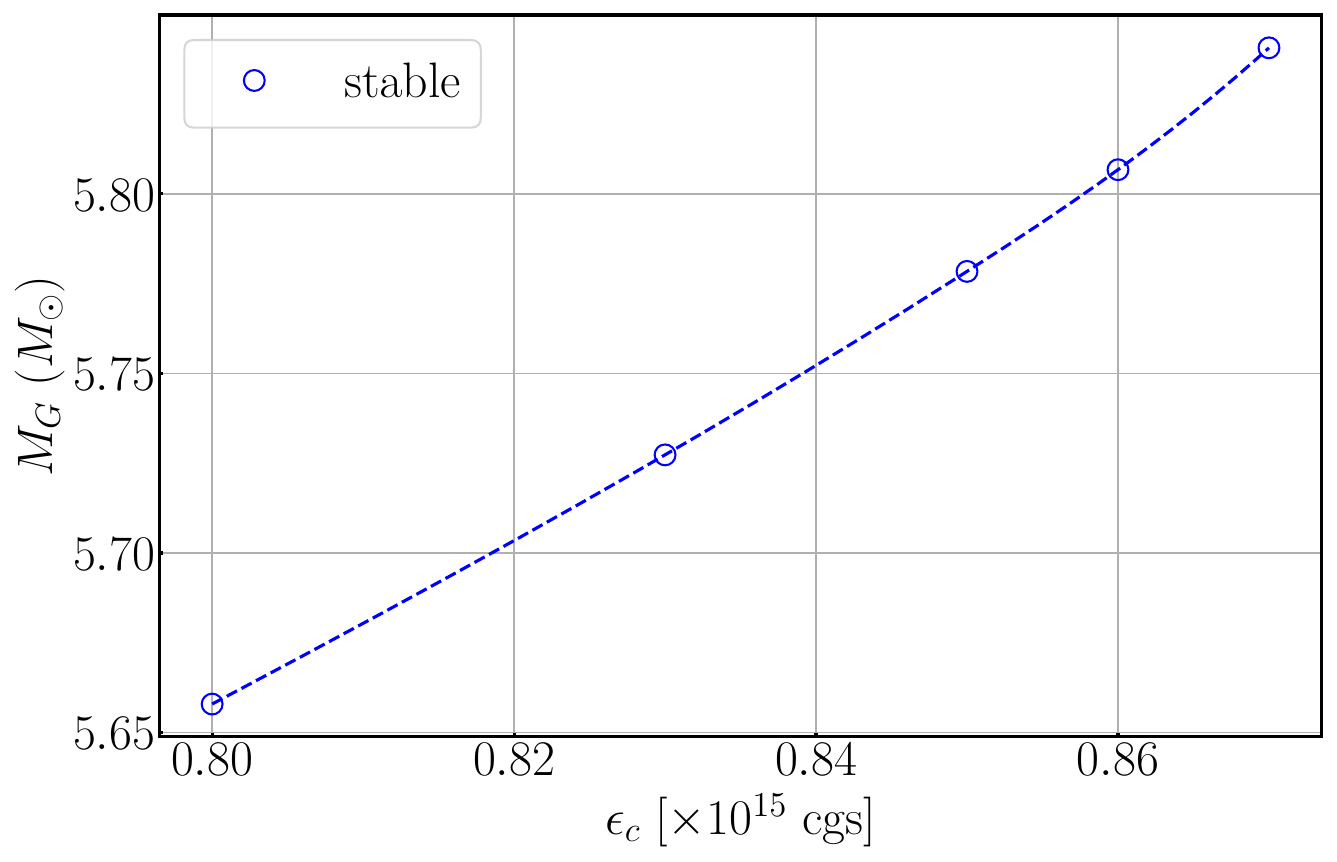}
        \caption{\texttt{m0.01\_B12\_J30}}
        \label{m0.01_J30}
    \end{subfigure}
    \caption{Dynamical stability of sequences (a) \texttt{m0.01\_B12\_J8} (b) \texttt{m0.01\_B12\_J12} (c) \texttt{m0.01\_B12\_J20} (d) \texttt{m0.01\_B12\_J40}. The blue circles and red crosses indicate the regions where the star is dynamically stable and unstable, respectively.}
    \label{fig:massive}
\end{figure*}

The chosen models are presented in \cref{fig:massive}. 
No turning point is found for the sequences with $J=20$ and 30 due to the same reason explained above, and the models are stable against axisymmetric perturbations.
For the sequences with $J=8$ and 12, a turning point exists and we find that the onset of instability is in the close vicinity of the turning point, i.e., the turning point criterion approximately holds.
In the massive theory, we also demonstrate that unstable models will collapse into a black hole within a dynamical timescale.
As a representative example, we plot the evolution of the maximum rest-mass density as well as the scalar field extracted at a certain distance inside the star for \texttt{m0.01\_B12\_J12} (red curves in \cref{fig:massless,fig:massive}).
We again observe a runaway growth in $\rho_{\rm max}$ and a strong suppression in the scalar profile after the black hole forms.
Following the collapse, the scalar field decays to a magnitude of $\sim10^{-3}$ over the dynamical timescale. 
The decay rate is much slower than the massless case at late times, as shown in the bottom panel of \cref{fig:unstable_evo}, and can be attributed to the dispersion relation of scalar waves \cite{Sperhake:2017itk,Cheong:2018gzn,Rosca-Mead:2020ehn,Geng:2020slq,Kuan:2022oxs}.
In particular, the propagation group speed of waves at the frequency $\omega_{\phi}$ is given as [Eq.~(27) in \cite{kuan23}]
\begin{align}
    v_g = (1+m_\phi^2 \lambdabar^2 )^{-1/2},
\end{align}
where $\lambdabar$ denotes the wavelength. 
It can thus be seen that the scalar waves with wavelengths $\lambdabar\gtrsim1/m_\phi$ (i.e., $\omega_\phi<m_\phi$) will dissipate over a prolonged damping timescale.

\begin{figure}
    \centering
    \includegraphics[width=\linewidth]{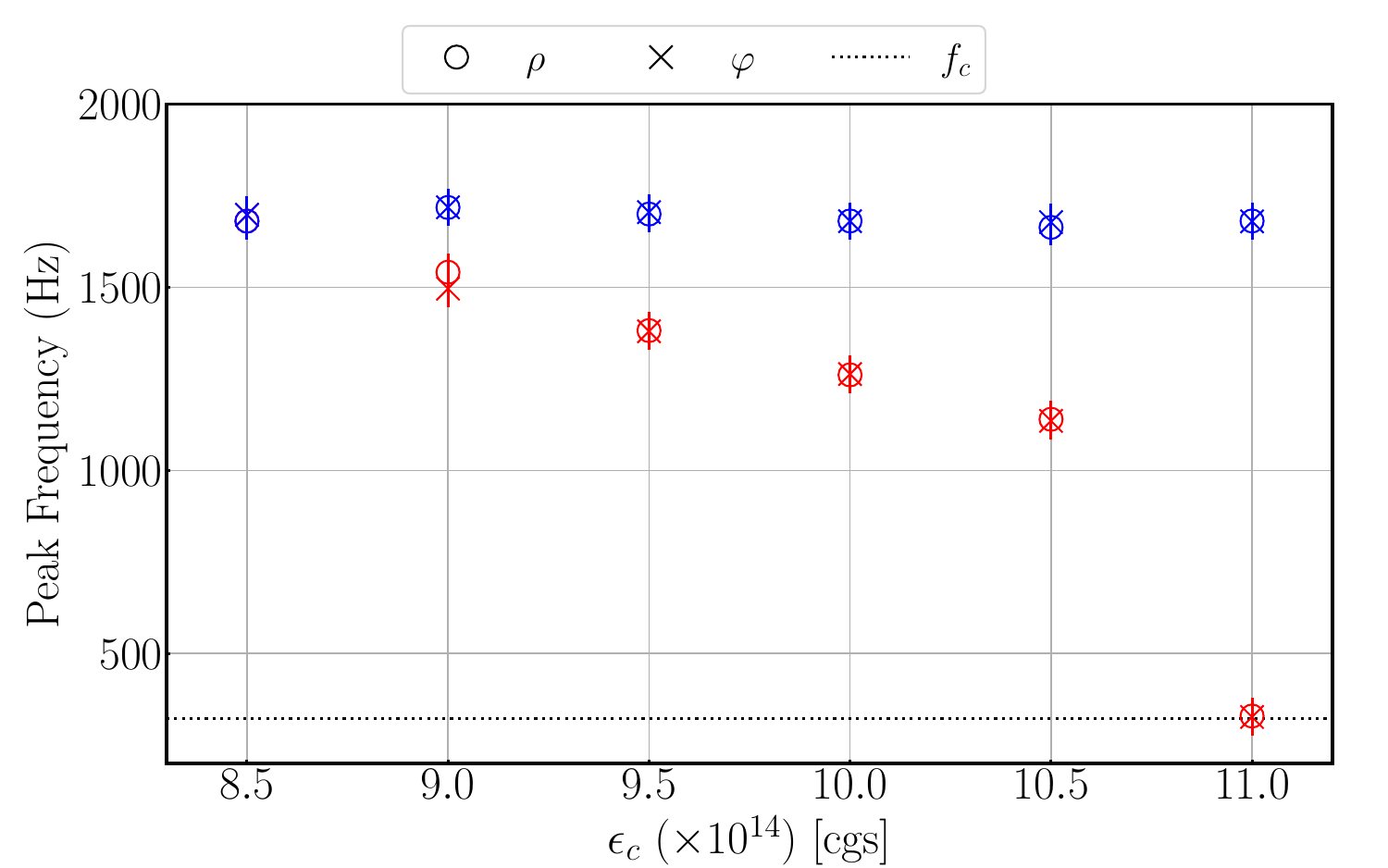}
    \caption{Spectrum of the axisymmetric oscillations in the frequency band $\le2$~kHz for the sequence \texttt{m0.01\_B12\_J8} (cf.~\cref{fig:massive} for its $M_G$-$\epsilon_c$ relation). Two classes of modes are observed, namely, the quasi-radial $m=0$ fundamental (blue) and $\phi$- (red) modes. The modes identified by the Fourier spectrum of oscillating central rest-mass density are marked as circles, while those identified from oscillating central scalar field are denoted as crosses. Agreement between the analysis of either quantity is observed. The Yukawa cutoff $f_c := \omega_\phi / (2 \pi)$ is presented as the dashed horizontal line.
    }
    \label{fig:mode_m0.01_J8}
\end{figure}

To further assess the turning point criteria,
we extract the spectrum of axisymmetric oscillations in the frequency band $\leq 2~{\rm kHz}$ for the sequence \texttt{m0.01\_B12\_J8} before the turning point as shown in \cref{fig:mode_m0.01_J8}.
The modes are extracted by performing Fourier analysis on the central rest-mass density $\rho_c$ and the central scalar field $\varphi_c$ marked as circles and crosses, respectively, in \cref{fig:mode_m0.01_J8}.
We observe that two classes of modes emerge in the spectrum,
which is speculated to be the quasi-radial $m=0$ fundamental mode (blue) and $\phi$-mode (red).
We found that the frequency of $\phi$-mode decreases as the central internal energy $\epsilon_c$ approaches the turning point
and eventually reaches a value very close to the Yukawa cutoff frequency of the scalar field $f_c:= \omega_\phi / (2\pi)$ in the stable model closest to the turning point.
This suggests that the dynamical instability near the turning point arises from the $\phi$-mode reaching the cutoff frequency, whereas, in GR, it is triggered by the fundamental mode hitting zero frequency (e.g., \cite{Kokkotas:2000up}). This is similar to the mode analysis in the static case \cite{Mendes:2018qwo,Blazquez-Salcedo:2021exm,Mendes:2021fon}.

%%%%%%%%%%%%%%%%%%%%%%%%%%%%%%%%%%
\section{Discussion}\label{discussion}
%%%%%%%%%%%%%%%%%%%%%%%%%%%%%%%%%%
By performing fully relativistic 2D simulations, we examine the well-known turning-point criterion dictating the condition for one kind of instability among many others. 
This criterion has been rigorously proven for rigidly rotating configurations by \citet{frie88} in pure GR, while the extension of it to more general configurations seems only plausible by the use of numerical simulations.
In this work, we evolve scalarized neutron stars along constant-angular-momentum sequences to pin down the onset of an axis-symmetric instability for various of the theory parameters as well as $J$ (\cref{tab:models}).
Our results suggest that the criterion for rigidly rotating bodies in GR [i.e., \cref{eq:turn1}] is largely valid also for differentially rotating stars in the DEF theory, and the observed onset of instability agrees within the numerical error at the turning point along the constant--$J$ sequences (\cref{fig:massless,fig:massive}).
For a representative stable model with $J=20$, we see that the density and scalar profiles as well as the rotational law are perfectly preserved when we terminate the simulation at $\gtrsim15$~ms (\cref{fig:snapshot}).
A word of caution is appropriate here.
Other instabilities, such as one-arm and bar-mode instabilities \cite{pasc17}, can be activated in reality as 3D simulations suggest \cite{shib05a,espi19}.
Here, the results are limited to axisymmetric (in)stability that is cared of in the turning-point criterion.

%%%%%%%%%%%%%%%%%%%%%%%%%%%%%%%%%%
\section*{Acknowledgement}
Numerical computation was performed on the clusters Sakura at the Max Planck Computing and Data Facility.
K.S. and S.Y.  are supported by the European Union-NextGenerationEU, through the National Recovery and Resilience Plan of the Republic of Bulgaria, project No. BG-RRP-2.004-0008-C01. DD  acknowledges financial support via an Emmy Noether Research Group funded by the German Research Foundation (DFG) under grant no. DO 1771/1-1. We acknowledge Discoverer PetaSC and EuroHPC JU for awarding this project access to Discoverer supercomputer resources.

%%%%%%%%%%%%%%%%%%%%%%%%%%%%%%%%%%%%%%%%

\appendix

\bibliography{references}

\end{document}